\documentstyle[12pt,aasms4,psfig]{article}

\slugcomment{to appear in the Astrophysical Journal (Letters)} 
 
\begin{document}
 
\title{Detection of CO (2--1) and Radio Continuum Emission from the
$z =$ 4.4 QSO BRI~1335$-$0417} 
 
\author{C. L. Carilli}
\affil{National Radio Astronomy Observatory, P.O. Box O, Socorro, NM,
87801, USA \\} 
\authoremail{ccarilli@nrao.edu}
\author{Karl M. Menten}
\affil{Max-Planck-Institut f{\"u}r Radioastronomie, Auf dem H{\"u}gel 69,
Bonn, D-53121, Germany \\}
\authoremail{kmenten@mpifr-bonn.mpg.de}
\author{Min S. Yun}
\affil{National Radio Astronomy Observatory, P.O. Box O, Socorro, NM,
87801, USA \\}
\authoremail{myun@nrao.edu}
 
\begin{abstract}

We have detected redshifted CO (2--1) emission at 43 GHz and radio
continuum emission at 1.47 and 4.86 GHz from the $z$ = 4.4 QSO
BRI~1335$-$0417 using the Very Large Array.  The CO data imply
optically thick emission from warm ($>30$ K) molecular gas with a
total mass, $M$(H$_2$), of 1.5$\pm$0.3 $\times10^{11}$ $M_\odot$, using
the standard Galactic gas mass-to-CO luminosity conversion factor.  We
set an upper limit to the CO source size of $1{\rlap.}{''}1$, and a
lower limit of $0{\rlap.}{''}23\times({{T_{\rm ex}}\over{50
K}}$)$^{-{1\over2}}$, where $T_{\rm ex}$ is the gas excitation
temperature. We derive an upper limit to the dynamical mass of
2$\times$10$^{10}\times$sin$^{-2}i$ $M_\odot$, where $i$ is the disk
inclination angle.  To reconcile the gas mass with the dynamical mass
requires either a nearly face-on disk ($i$ $<$ 25$^\circ$), or a gas
mass-to-CO luminosity conversion factor significantly lower than the
Galactic value.  The spectral energy distribution from the radio to
the rest-frame infrared of BRI~1335$-$0417 is consistent with that
expected from a nuclear starburst galaxy, with an implied massive star
formation rate of 2300$\pm$600 $M_\odot$ yr$^{-1}$.

\end{abstract}
 
\keywords{radio continuum: galaxies --- infrared: galaxies ---
galaxies: distances and redshifts, starburst, evolution, radio lines} 

\section {Introduction}

By selecting for very red, point-like optical sources, McMahon (1991) 
has identified a large sample of $z \ge 4$ QSOs from the APM survey 
(see  Irwin, McMahon, \&\ Hazard 1991;
Storrie-Lombardi et al. 1996).  Omont et al. (1996a) have made
the remarkable discovery that a significant fraction of these high $z$ QSOs
show thermal dust emission.  Of their sample of 16 QSOs at $z > 4$, a total of
6 sources
show dust emission at 230 GHz, with implied dust masses
of 10$^8$ to  10$^9$ $M_\odot$. Follow-up observations of
three of these 
dust emitting QSOs have revealed CO emission with implied molecular
gas masses $\sim$ 10$^{11}$ $M_\odot$ (Ohta
et al. 1996; Omont et al. 1996b; Guilloteau et al. 1997, 1999).
Omont et al. (1996a) speculate that ``$\ldots$such large amounts of
dust [and gas] 
imply giant starbursts at $z > 4$, at least comparable to those found
in the most hyperluminous IRAS galaxies$\ldots$''.  However, the evidence
for active star formation in these sources remains circumstantial,
primarily based on the presence of large gas reservoirs. It remains
possible that the dust is heated by the active galactic nucleus (AGN),
rather than by a starburst. 

The source BRI~1335$-$0417, which has an optical redshift of $4.396
\pm 0.026$, is the second brightest source at millimeter wavelengths
in the Omont et al. (1996a) sample.  Its optical spectrum shows strong
absorption by Ly $\alpha$ and metal lines at the source redshift
(Storrie-Lombardi et al. 1996).  Guilloteau et al. (1997) have
detected CO (5--4) line emission toward BRI~1335$-$0417 with an
implied gas mass $\sim$ 10$^{11}$ $M_\odot$.  The CO(5--4) line has a
width of 420 $\pm$ 60 km s$^{-1}$ (FWHM) and is centered at $z =
4.4074 \pm 0.0015$.

In this paper we present detections of redshifted CO (2--1) emission
at 43 GHz and radio continuum emission at 1.47 GHz and 4.86 GHz using
the Very Large Array (VLA). These data constrain the temperature of
the molecular gas to be fairly high ($>$ 30 K). We find that the shape
of the spectral energy distribution (SED) of BRI~1335$-$0417 from the
radio into the rest frame infrared is similar to that of the
nuclear starburst galaxy M82. Our data support the idea that the radio
continuum and thermal dust emission are powered by a starburst that is
concurrent with the AGN activity in BRI~1335$-$0417. We use $H_0$ = 75
km s$^{-1}$ Mpc$^{-1}$ and $q_0$ = 0.5.

\section{Observations and Results}

BRI~1335$-$0417 was observed with the 3 km (C) configuration of the
NRAO\footnote{The National Radio Astronomy Observatory (NRAO) is operated
by Associated Universities, Inc. under a cooperative agreement with the
National Science Foundation.} VLA
at 4.860 GHz on 1999 Feb. 21 and 26 for a total of 4 hours and at 1.465
GHz on Nov. 5 and Dec. 26, 1999 for a total of 2 hours.  Standard
amplitude and phase calibration were applied, and the absolute flux
density scale was set by observing 3C~286. Standard
self-calibration and high dynamic range imaging procedures were
employed (Perley 1999).  The response of the ``dirty'' beam was
deconvolved using the CLEAN algorithm to a minimum residual surface
brightness level of 2 times the $1\sigma$ rms noise in 
the images.   The 4.86 GHz image was restored with a Gaussian beam
of size $18''\times7''$ (FWHM) and a major axis position angle (PA)
of 114$^\circ$, measured east of north. The restoring beam of the 1.47 GHz
image was a 18$''\times17''$ Gaussian of  PA  44$^\circ$.

The source was detected at both 1.47 GHz and 4.86 GHz. 
It appears unresolved at both frequencies and
its flux density, $S_\nu$, is 220$\pm$43 $\mu$Jy
and 76$\pm$11 $\mu$Jy at 1.47 and 4.86 GHz, respectively.
The source position at 4.86 GHz is
R.A.= $13^{\rm h}38^{\rm m}03{\rlap.}{^{\rm s}}50$,
decl.= $-$04$^\circ32'35{\rlap.}{''}5$ (J2000)
with an uncertainty of
$\pm 0{\rlap.}{''}3$ in both coordinates.  This is
consistent, within the errors, with the positions of BRI~1335$-$0417's
optical, mm continuum, and CO (5--4) line emission. The spectral
index, $\alpha$,  between 1.47 and 4.86 GHz is $-$0.82$\pm$0.17,
where $S_\nu$ $\propto$ $\nu^{\alpha}$.

We observed BRI~1335$-$0417 with the 1 km (D) configuration of the VLA at
42.6~GHz on 1999 March 7 and 9 for a total of 16 hours. The absolute
flux density scale was set using 3C~286 for which we assumed a flux density
of 1.5 Jy corrected for atmospheric
opacity to the top of the atmosphere. Fast switching phase calibration
was employed (Carilli \& Holdaway 1999). We switched with a total cycle time of
180~s between BRI~1335--0417 and the celestial calibration source
1334$-$127.  The rms phase variations after calibration were
$\le$ 15$^\circ$ on the longest (1 km) baselines. The spatial resolution is
$2{\rlap.}{''}3\times1{\rlap.}{''}7$ (FWHM, major axis oriented north-south).

A severe limitation at the VLA for observing broad lines at high
frequencies is the maximum correlator bandwidth of 50 MHz, and the
limited number of spectral channels (7) when using this bandwidth and
dual polarization. The bandwidth of 50 MHz corresponds to a velocity
coverage of only 350 km s$^{-1}$ at 43 GHz, which is further reduced
by bandpass roll-off at the band edges. While this is comparable to
the CO (5--4) line FWHM of 420 $\pm$ 60 km s$^{-1}$, some emission in
the line wings will be lost.  Since these correlator limitations
preclude a meaningful determination of the line profile, we chose to
observe in continuum mode, using 2 IFs with 2 polarizations each. One
of the IFs was centered on the emission line (42.635 GHz), while the
second was centered 1400 km s$^{-1}$ off the line (42.835 GHz). This
observing set-up maximizes sensitivity to the integrated line
intensity but does not yield information on the line profile, which
was well-determined for the CO (5--4) line by the observations of
Guilloteau et al. (1997).  Assuming that the CO (2--1) and (5--4) line
profiles are similar, and taking into account the limited width of the
continuum band employed, we estimate that we are missing about 37$\%$
of the velocity-integrated line emission in BRI~1335$-$0417.

The VLA images (Fig. 1) show a clear detection of CO (2--1) emission
from BRI~1335$-$0417 on both observing days.  Fig. 1 shows the on-line
and off-line images made from the data taken on each day separately,
as well as from the averaged dataset.  From the map made from the
averaged data, we estimate a peak flux density of 0.74 $\pm$ 0.12 mJy
at R.A.= $13^{\rm h}38^{\rm m}03{\rlap.}{^{\rm s}}41$, decl.=
$-$04$^\circ32'34{\rlap.}{''}8$ (J2000) with an uncertainty of $\pm
0{\rlap.}{''}3$ in both coordinates.  No emission is seen in the off
image at the position of the quasar, to a 2$\sigma$ upper limit of
0.24 mJy.  The CO source's observed brightness temperature is 0.12 K.
The CO emission region is spatially unresolved, with an upper limit to
its size of $1{\rlap.}{''}1$ determined from Gaussian model fitting.

The velocity-integrated line intensity for the CO (2--1) line from
BRI~1335$-$0417 is 0.32$\pm$0.06 Jy km s$^{-1}$, corrected for the 37$\%$
missing from the VLA bandpass.
Using the CO (5--4)
data of Guilloteau et al. (1997), we determine a
velocity-integrated CO (2--1)-to-(5--4)
line flux ratio of 8.7$\pm$1.9. This
value is roughly  consistent with the constant
brightness temperature value of 6.25 (i.e. integrated line flux
increasing as $\nu^2$) expected for optically thick emission in both
lines, for a fixed  source size. 

Extrapolating the dust continuum spectrum 
from 230 GHz, where 8 mJy are observed (Guilloteau et al. 1997),
to 43 GHz using a spectral index
of +3.25, one calculates an expected continuum flux density of 
34 $\mu$Jy at the latter frequency, which is well below
our noise level of 0.12 mJy in the off-line channel. 

\section{Discussion}

The critical question concerning BRI~1335$-$0417 is whether the dust
and radio continuum emission are powered by the AGN, or star
formation, or both.  Fig. 2 shows the integrated spectral energy
distribution for BRI~1335$-$0417 for rest frame frequencies from 7.6
GHz up to 4600 GHz, normalized to the spectrum of the nuclear
starburst galaxy M82. For demonstrative purposes we have quantified
the spectrum of M82 using two accurate polynomial fits: one to the
synchrotron emission which dominates at frequencies below 100 GHz, and
a second to the thermal dust emission which dominates above 100
GHz. All the data points for BRI~1335$-$0417 fall within 
2$\sigma$ of the M82 curve.  The agreement between the SED of
BRI~1335$-$0417 and M82 argues in favor of star formation playing the
dominant role in heating the dust and powering the radio continuum
emission from BRI~1335$-$0417,  although we
cannot preclude a contribution from the AGN.

It is important to note that, while
the shapes of the SEDs are similar, the luminosity of BRI~1335$-$0417
is 1.2$\times10^{32}$ ergs s$^{-1}$ Hz$^{-1}$ at 1.47 GHz in the rest
frame of the source, assuming a radio spectral index of $-0.8$, while
that of M82 is a factor 1400 lower (assuming a distance of 3 Mpc to
M82).  For comparison, the 
rest frame 1.47 GHz luminosity of BRI~1335$-$0417 is a factor 40 larger
than that of Arp 220, and even a factor 2 larger than that of the
nearby low luminosity radio galaxy M87.
If the emission is powered by star formation, 
a very  rough estimate of the massive star formation rate in BRI~1335$-$0417 
can be derived from the radio and submm continuum emission using
equations 1 and 2 in Carilli \& Yun (1999). Both values are consistent
with a star formation rate of 2300$\pm$600 $M_\odot$ yr$^{-1}$. We
emphasize that this is at best an order-of-magnitude estimate.

That the CO (2--1) and (5--4) lines from BRI~1335$-$0417 have, within the
uncertainties, equal brightness temperature implies that the
molecular gas must be fairly warm, with a lower limit to the
rotational temperature of 30 K set by the excitation of the CO (5--4)
line, and that the line emission is optically thick.  Our VLA imaging
implies an upper limit to the CO source size of $1{\rlap.}{''}1$.  A lower
limit to the source size can be derived assuming a single,
homogeneous, optically thick emitting region (Ohta et al. 1996). The
beam filling factor is given by: ${\Omega_{\rm s}}\over{\Omega_{\rm
b}}$ = ${{T_{\rm obs}}\over{T_{\rm ex}}} \times (1+z)$, where
$\Omega_{\rm s}$ and $\Omega_{\rm b}$ are the source and 
beam solid angles, respectively, $T_{\rm obs}$ is the observed line brightness
temperature (0.12 K), and the line excitation temperature, $T_{\rm ex}$, is
assumed to be equal to the intrinsic (rest frame)
source brightness temperature.  This calculation leads to a minimum
source diameter of
$0{\rlap.}{''}23\times({{T_{\rm ex}}\over{50 K}})^{-{1\over2}}$.

An interesting problem that has arisen in the study of extreme nuclear
starburst galaxies is that the implied gas mass is comparable to, or
sometimes greater than, the dynamical mass (Tacconi et al. 1999;
Bryant \&\ Scoville 1996; Ohta et al. 1996; Scoville, Bryant, \&\ Yun
1997; Downes \&\ Solomon 1998).  BRI 1335--0417 is an extreme example
of this ``excess mass'' problem.  From the CO(2-1) luminosity we
estimate a molecular gas mass, 
$M$(H$_2$), of 1.5$\pm$0.3 $\times10^{11}$, using
equation 5 in Solomon, Radford, and Downes (1992), which
assumes the standard Galactic gas mass-to-CO luminosity conversion
factor. We have also included an upward
correction to the mass of 30$\%$ arising from the fact that the 
line brightness temperature is measured relative to that of the
microwave background radiation. This correction factor is 30$\%$ for a
gas at 50 K at z = 4.4. Using the upper limit to the source
radius of $0{\rlap.}{''}55$ or 2.25 kpc, and assuming a rotational velocity of
200 sin$^{-1}i$ km~s$^{-1}$, where $i$ is the disk inclination
angle with respect to the sky plane, leads to an upper limit to the
dynamical mass of $2\times10^{10}\times$ sin$^{-2}i~M_\odot$.
Comparing this to the gas mass 
leads to an upper limit to the inclination angle of $i$ $<$
$25^\circ$.  The fact that the source is a broad line QSO is consistent
with a disk rotation axis oriented within $45^\circ$ of our line of sight
(Barthel \&\ Miley 1989). However, the persistence of this mass
problem for a number of sources is becoming difficult to reconcile
with all the sources being close to face-on.

For high redshift
sources, gravitational lensing could act to magnify the CO emission,
leading to an  overestimate of the gas mass. However,  
for BRI 1335--0417 there is thus far no evidence for
multiple imaging due strong gravitational lensing. Another way to
mitigate the excess mass problem is to adjust the cosmological
parameters.  However this would require a decrease in the luminosity
distance, i.e. an increase in $H_0$ or $q_0$, which would be
contrary to current observational evidence (Peebles 1999; Turner 1999).

Perhaps the most likely cause for the excess mass problem is that
applying the Galactic gas mass-to-CO luminosity conversion factor
causes an overestimate of the gas mass for extreme objects such as BRI
1335--0417.  In particular, it has been shown that assumptions
underlying the standard Galactic conversion factor involving CO
emission from a conglomeration of many small, highly optically thick
clouds lead to overestimates of the gas mass by a factor 3 or more
relative to a single homogeneous disk model (Combes, Maoli, \&\ Omont
1999; Downes \&\ Solomon 1998; Solomon et al. 1997).  Assuming a
factor 3 decrease in the gas mass leads to an upper limit to the
inclination angle, $i$, of $48^\circ$. Overall, a combination of small
inclination angle plus a lower gas mass-to-CO luminosity conversion
factor may be sufficient to solve the
excess mass problem for BRI 1335--0417. Nevertheless, it appears likely that
in this source the dynamical mass is dominated by the molecular gas
mass on scales of a few kpc.

BRI~1335$-$0417 has similar properties to the well-studied QSO
BR~1202$-$0725 at $z = 4.9$, including (i) similarity of the SED with
that of the starburst galaxy M82 (Kawabe et al. 1999; Yun et
al. 1999), (ii) similar 
dust and gas masses (Ohta et al. 1996; Omont et al. 1996b), (iii) a
similar massive star formation rate as derived from the radio
continuum and dust emission (Kawabe et al. 1999), (iv) a constant
brightness temperature for the CO emission,
at least up to the (5--4) transition (Kawabe et al. 1999), and (v)
similar optical spectra, in particular, strong associated absorption
lines suggesting a gas-rich environment (Storrie-Lombardi et al. 1996).
Kawabe et al. (1999) have made a detailed comparison between 
the CO emission from BR~1202$-$0725 and that
from the starburst nucleus in M82. The nucleus of M82 also shows
constant brightness temperature up to CO (5--4)
(G{\"u}sten et al. 1993). From this comparison, Kawabe et al. conclude that
at least half of the molecular gas mass has a temperature
between 50 K and 70 K and a 
density between 10$^4$ cm$^{-3}$ and 10$^5$ cm$^{-3}$. The pressure in
the molecular gas is then a few times $10^{-10}$ dyn cm$^{-2}$. Such
high pressures have been proposed for massive nuclear starburst
regions in nearby ultraluminous infrared galaxies (Condon et
al. 1991).

A significant difference between BRI~1335$-$0417 and BR~1202$-$0725 is that
the latter is a double source with a separation of 4$''$ in the cm
and mm continuum as well as in CO line emission (Omont et al. 1996b;
Yun et al. 1999), 
while BRI~1335$-$0417 appears unresolved at  arcsecond resolution.
It has been argued that the double structure of BR~1202$-$0725 is caused by
gravitational lensing, which might lead to an
overestimate of this source's intrinsic dust and gas masses and star
formation rate by an order of magnitude (Omont et al. 1996b).
However, the lensing hypothesis is complicated by the fact that the
optical QSO is a single source (Hu, McMahon, \&\ Egami 1996).
If BRI~1335$-$0417 is gravitationally lensed, the multiple images must be
separated by less than 1$''$.

Tan, Silk, \&\ Blanard (1999) have presented a model of the star
formation history of the universe which may explain the faint source
counts from the optical through the radio regimes. They hypothesize
that star formation occurs in two phases. At high redshift ($2 < z <
5$), very active star formation occurs in merger events leading to the
formation of the spheroidal components of galaxies (spiral bulges and
ellipticals). At lower redshift ($ z< 2$), the stellar disks of
galaxies are formed over longer periods at lower star formation rates.
The high $z$ starbursts are dusty systems and are revealed primarily
as constituents of the faint submm, IR, and radio source populations.
In contrast, the faint optical source counts are dominated by the
lower $z$ disk-forming systems.  Tan et al. state that the early
spheriods may act as ``seeds'' for disk formation as gas falls in, and
that the present-day stellar mass in bulges and halos is comparable to
that in disks. Our observations of BRI~1335$-$0417 are consistent with
this source being a member of the high $z$ starburst population.
These data support the speculation by Omont et al. (1996b) that:
``...the emergence of quasars at very high redshift is connected with
the onset of galaxy formation, possibly with the formation of the
cores of elliptical galaxies''.

Overall, the cm and mm continuum observations, as well as the CO line
data, argue in favor of a very massive starburst concurrent with the
AGN activity in BRI~1335$-$0417 at $z = 4.4074$.  The gas depletion
timescale is short, $\le$ 10$^9$ years, in which time a significant
fraction of the stars in the host galaxy of the AGN may be formed 
(Benford et al. 1999).
Further observations are required to understand the nature of
BRI~1335$-$0417, including: (i) high resolution imaging with the VLA
of the radio continuum emission to search for multiple imaging due to
gravitational lensing on a scale of a few hundred mas, and/or to
search for evidence of a radio jet, (ii) VLA imaging at 100 mas
resolution of the CO emission to resolve the optically thick line
emitting regions, and (iii) observations of higher order CO
transitions to constrain the temperature of the molecular gas.

\vskip 0.2truein 

We would like to thank F. Owen, D. Downes, M. Walmsley, J. Lehar, and
the referee, A. Omont, for useful comments and
discussions. We would also like to thank J. van Gorkom for her
cooperation in dynamic scheduling of the VLA. 
C.C. acknowledges support from the Alexander von Humboldt Society. 
This research made use of the NASA/IPAC Extragalactic
Data Base (NED) which is operated by the Jet propulsion Lab, Caltech,
under contract with NASA.


\newpage

\begin{figure}
\vskip -1in
\psfig{figure=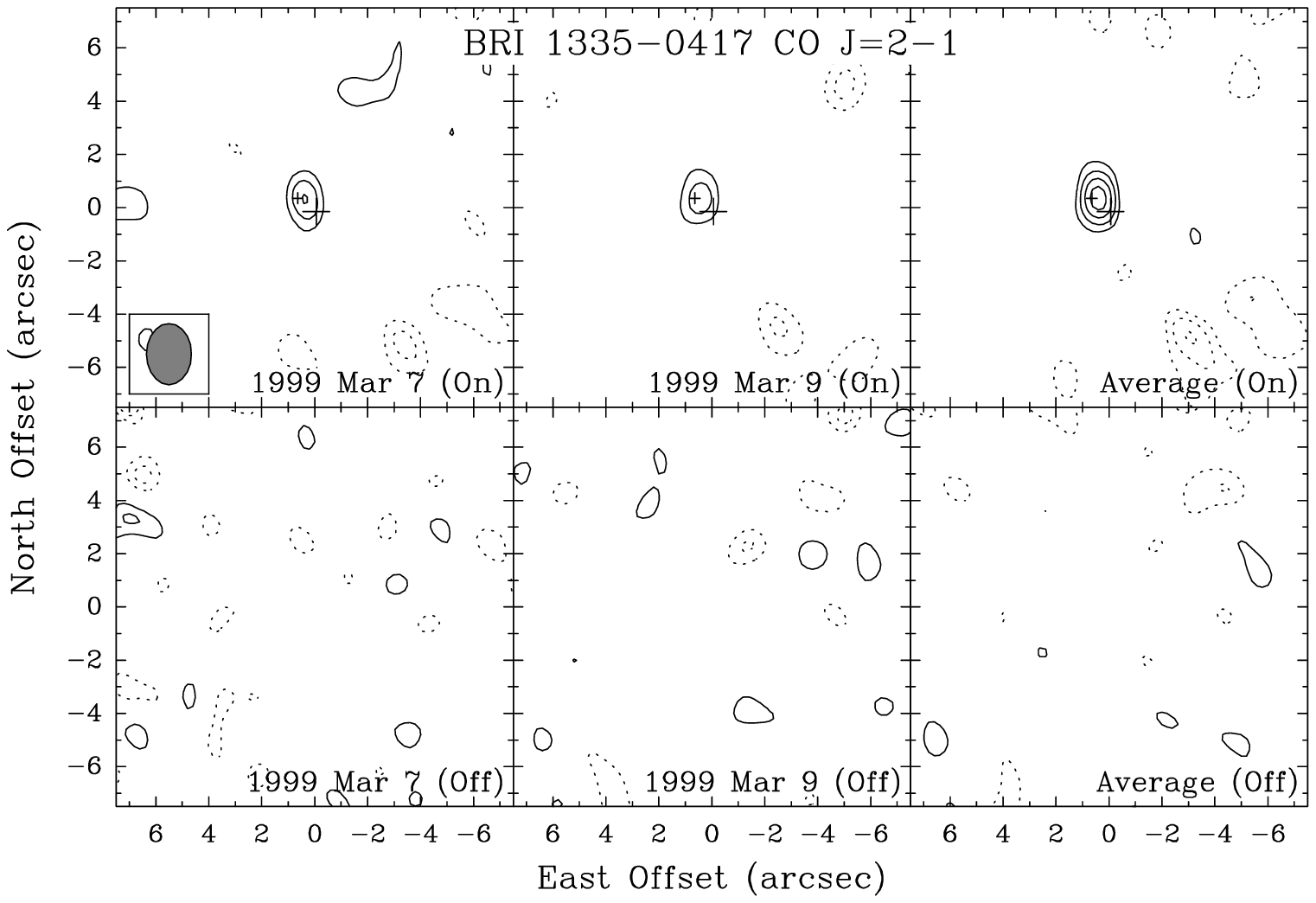,width=7in}
\vspace*{-4in}
\caption{VLA contour maps of BRI~1335$-$0417 taken at a
frequency centered on (upper panels), and off (lower panels) the
frequency of the redshifted ($z = 4.4074$) CO (2--1) line emission
near 43~GHz.  The leftmost and middle panels show the maps produced
from data taken on 1999 March 7 and 9, respectively, while the
rightmost map was produced from the average of these individual
datasets.  The contour levels represent $-$4, $-$3, $-$2, 2, 3, 4, and
5 times the $1\sigma$ rms noise, which is 0.19 mJy beam$^{-1}$ for the
``single day'' maps and 0.13 mJy beam$^{-1}$ for the map made from the
averaged data.  The maps have been restored with a beam of size
$2{\rlap.}{''}3\times1{\rlap.}{''}7$
(FWHM) elongated N-S, which is indicated by the
grey ellipse in the leftmost ``on'' map.  The small cross marks the
position of the 1.35 mm dust continuum source determined by Guilloteau
et al. 1997, with the size of the cross indicating the $\pm
0{\rlap.}{''}2$ uncertainty of the mm position.  The larger cross
marks the optical position given by Storrie-Lombardi et al.  1996
without errors quoted. }
\end{figure}
\vfill\eject

\begin{figure}
\vskip -2in
\hspace*{-1.7in}
\psfig{figure=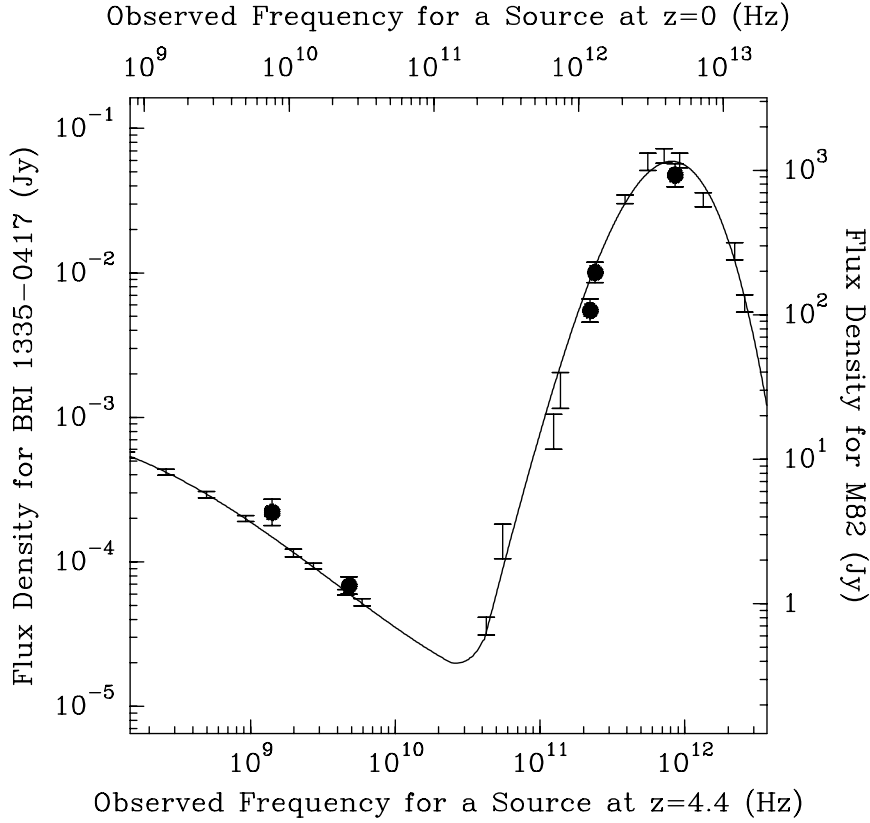,width=10in}
\vspace*{-8.0 in}
\caption{
The solid line shows the radio-through-infrared
spectral energy distribution of M82 derived via polynomial fits to the
observed data points, which are taken from the NASA Extragalactic Data 
Base (NED) and are shown as vertical error bars. 
The solid dots (plus error bars) show the data for the
$z = 4.4074$ QSO BRI~1335$-$0417 (Omont et al. 1996a; Guilloteau et al.
1997; Benford et al. 1999; this paper). For comparative purposes, the data for
BRI~1335$-$0417 have been
normalized to the SED of M82 using the mean of the normalization value
at rest frame frequencies of 26 GHz and 1300 GHz.
The left and right hand flux density scales are those appropriate for
BRI~1335$-$0417 and M82, respectively. 
Note that the intrinsic luminosity at 1.4 GHz of BRI~1335$-$0417 is a
factor 1400 times larger than that of M82.
The lower abscissa is chosen to represent the observed frequency scale for
the  BRI~1335$-$0417 data points  (ie. for a source at z = 4.4),
whereas the upper abscissa is appropriate
for the M82 data points (ie. for a source at z = 0). 
}
\end{figure}

\vfill\eject

\end{document}